# Graphene-Wrapped Sulfur Particles as a Rechargeable Lithium-Sulfur-Battery Cathode Material with High Capacity and Cycling Stability

Hailiang Wang,[1,§] Yuan Yang,[2,§] Yongye Liang,[1] Joshua Tucker Robinson,[1] Yanguang Li,[1] Ariel Jackson,[2] Yi Cui[*,2] and Hongjie Dai[*,1]

[1]*Department of Chemistry, and* [2]*Department of Materials Science and Engineering, Stanford University, Stanford, CA 94305, USA.*

[§] These two authors contributed equally to this work

*\* Correspondence to* yicui@stanford.edu *and* hdai@stanford.edu

## Abstract

We report the synthesis of a graphene-sulfur composite material by wrapping polyethyleneglycol (PEG) coated submicron sulfur particles with mildly oxidized graphene oxide sheets decorated by carbon black nanoparticles. The PEG and graphene coating layers are important to accommodating volume expansion of the coated sulfur particles during discharge, trapping soluble polysulfide intermediates and rendering the sulfur particles electrically conducting. The resulting graphene-sulfur composite showed high and stable specific capacities up to ~600mAh/g over more than 100 cycles, representing a promising cathode material for rechargeable lithium batteries with high energy density.

There has been a steady increase in demand for clean and efficient energy storage devices due to the ever-rising concerns about limited global energy supply and environment and climate changes.[1-7] Due to high volume and gravimetric energy density, rechargeable lithium batteries have become the dominant power source for portable electronic devices including cell phones and laptops.[3-7] However, the energy and power densities of rechargeable lithium batteries require significant improvement in order to power electric vehicles[3-7] that are important to reducing fossil fuel consumption and $CO_2$ emission. Thus far, the lower specific capacities of cathode materials (~150mAh/g for layered oxides and ~170mAh/g for $LiFePO_4$) compared to those of the anode (370mAh/g for graphite and 4200mAh/g for Si) have been a limiting factor to the energy density of batteries. It is highly desirable to develop and optimize high capacity cathode materials for rechargeable lithium batteries.

Sulfur is a promising cathode material with a theoretical specific capacity of 1672mAh/g,[7-17] ~5 times higher than those of traditional cathode materials based on transition metal oxides or phosphates. Sulfur also possesses other advantages such as low cost and environmental benignity. Nevertheless, it has been difficult to develop a practical Li-S battery partly limited by the problems of low electrical conductivity of sulfur, dissolution of polysulfides in electrolyte and volume expansion of sulfur during discharge. These problems cause poor cycle life, low specific capacity and low energy efficiency.[7-17] Various carbon/sulfur composites utilizing active carbon, carbon nanotubes or mesoporous carbon have been made with specific capacity exceeding 1000mAh/g achieved.[8-17] However, it remains challenging to retain high and stable capacity of sulfur cathodes over more than 100 cycles.

Graphene is a two dimensional one-atom-thick conductor with high surface area, chemical stability, mechanical strength and flexibility, making it a useful growth substrate to anchor active materials for electrochemical energy storage applications.[18-26] Recently we have developed a two-step method to grow metal hydroxides and oxides nanocrystals with interesting morphologies and nanoscale sizes on graphene sheets with various degrees of oxidation and surface functional group coverage.[18-22] As a result of controlled nucleation and growth, the hydroxide or oxide nanomaterials are selectively formed on graphene with intimate interaction to the conducting graphene substrate. The strong electrical coupling renders the otherwise insulating active materials conducting, which significantly increases the specific capacitance (capacity) and rate capability of the graphene hybrid or composite electrode materials.

In principle, graphene-sulfur composite could lead to improved sulfur cathode materials for Li-S batteries. However, in addition to interfacing sulfur with graphene sheets, it is important to obtain sulfur particles well coated and confined by graphene sheets and meanwhile integrate polymeric [e.g., polyethyleneglycol (PEG)] 'cushions' in the hybrid structure.[12] These factors could minimize the dissolution and diffusion of polysulfides and accommodate volume expansion effects during discharge, therefore improving the cycling life of the sulfur cathode. One way to deposit sulfur onto graphene is by impregnating aggregated graphene sheets with melted sulfur to take advantage of the fact that sulfur interacts strongly with carbon.[8-17] This approach could improve the conductivity of sulfur but is not effective in containing the polysulfides. The pores in a graphene aggregate is irregular in shape and widely distributed in size due to the micron-scale size and sheet-like shape of individual graphene sheets. Therefore, polysulfides formed during discharge can still readily diffuse out of the graphene structure and initiate the "shuttle" problem which significantly undermines the cycling stability of the cell.[8,16,17] In a recent report by Wang et al, sulfur was melted and incorporated into graphene aggregates.[15] Even with a low loading of ~22 wt% sulfur in the graphene-sulfur composite, the cycling performance was not improved compared to pure sulfur.[15]

Here we present a rational design and synthesis of a novel graphene-sulfur composite material. PEG-containing surfactant coated sulfur particles are synthesized and wrapped by carbon black decorated graphene oxide sheets in a simple assembly process. With relatively stable specific capacity of ~600mAh/g and less than 15% decay over 100 cycles, the graphene-sulfur composite shows promising characteristics as a high performance cathode material for Li-S batteries.

In our approach (Figure 1), a mildly oxidized graphene oxide (mGO, with a lower degree of oxidation[21] than the commonly used Hummers GO[27]) material was used to form composites with sulfur particles. Carbon black nanoparticles (Super P, ~50nm in diameter, ~50% of mGO by mass) were first loaded onto mGO by simple bath sonication aimed at increasing the conductivity of the mGO sheets and the final composite material. The carbon black decorated mGO sheets were well dispersed in water since mGO sheets contained both hydrophobic aromatic regions to interact with carbon black and highly hydrophilic regions (oxygen functional groups) for dispersion in water.[28] Sulfur particles were synthesized by reacting sodium thiosulfate with hydrochloric acid in an aqueous solution of Triton X-100 (a surfactant with a PEG chain). The PEG surfactant was used as a capping agent for sulfur particles and to limit the size of sulfur particles to the sub-micron region during synthesis.[29] Small sulfur particle sizes should favor high specific capacity and rate capability of sulfur cathodes. The mGO/carbon black suspension was then mixed with surfactant coated sulfur particle to afford the final sulfur-graphene oxide composite.

Figure 2 shows scanning electron microscopy (SEM) images of the graphene-sulfur composite. The average size of the sulfur particles was found to be less than one micron, resulted from the size limiting effect afforded by the Triton-X100 stabilizing surfactant coating on the sulfur particles. A zoom-in SEM image (Figure 2b) showed graphene sheets coated around a sulfur particle. To verify the structure and composition of our graphene-sulfur composite, we carried out energy dispersive spectroscopic (EDS) mapping/imaging of our material. Chemical mapping confirmed that the bright particles in the SEM image (Figure 3a) were sulfur (Figure 3c, sulfur mapping), with overlaying C signals due to mGO coating on the sulfur particles (Figure 3d, carbon mapping). Colorimetric titration experiments[30] determined that the graphene-sulfur composite contained ~70 wt% of sulfur (with ~15% of mGO and ~8% of carbon black).

Coin cells were fabricated to test the electrochemical performance of our graphene-sulfur cathode material with a Li foil as the anode in an electrolyte of 1.0M lithium bis-trifluoromethanesulfonylimide in 1,3-dioxolane and 1,2-dimethoxyethane (volume ratio 1:1). The graphene-sulfur working electrodes was heated at 140°C for 5min to anneal the electrode and remove exposed sulfur that was not well coated by graphene (~10-20%). An initial discharge capacity higher than 1000mAh/g was observed. Figure 4a shows the 10[th] cycle charge and discharge curves of the composite material at different C rates (C rate was based on the theoretical specific capacity of sulfur, where a 1C rate corresponded to a current density of ~1673mA/g). The discharge curves exhibited multiple steps/stages corresponding to sequential reduction from S to $Li_2S$, while the charge process due to oxidation appeared relatively simple.[8] Cycling performance of the Li-S cell is shown in Figure 4b. At a rate of C/5, an initial capacity of ~750mAh/g was measured, followed by a decrease to a relatively stable capacity of ~600mAh/g after ~10 conditioning cycles. Within the next 90 cycles, the capacity only decreased by 13%, showing good cycling stability of the graphene-sulfur composite cathode. At a higher rate of C/2, even better cycling performance (9% of decay from the 10[th] to the 100[th] cycle) was achieved for the graphene-sulfur composite (Figure 4b), which was likely due to less shuttle effect at higher current density.

We have repeatedly observed high capacity and good cycling stability of our graphene-sulfur composite material, and the cycling stability was comparable to the best achieved cycling performance of sulfur cathode materials prepared by various methods. However, we found that the Li-S batteries generally showed much greater performance variations than other types of rechargeable lithium batteries. For example, about 30-50% of our Li-S cells showed worse cycling performance with ~20%-25% decay per 100 cycles. Such variability between cells was not understood currently and could be due to the complex cycling reaction processes in the Li-S battery and inhomogeneities in the film of the cathode material. Future work is needed to understand and reduce such variability.

A control sample of surfactant coated sulfur particles without graphene coating (but contained carbon black) showed much faster capacity decay and lower stable capacity (Figure 4c) than the one with graphene coating (Figure 4b). The specific capacity decreased from an initial ~700mAh/g to less than ~330mAh/g after 20 cycles (Figure 4c). In another control experiment, we found that graphene sheets mixed with sulfur particles without surfactant coating also failed to afford good cycling performance with continuous capacity decay over 50 cycles (Figure 4d). These results suggested that the graphene and PEG coating layers on sulfur particles are two important factors to the observed cycling stability of sulfur particles in our graphene-sulfur composite. The PEG containing surfactant coating on the sulfur particles could provide a flexible cushion in the sulfur-graphene composite to accommodate stress and volume changes. In addition, PEG chains have been suggested capable of trapping polysulfides.[12] Further, the carbon black decorated mGO sheets on the PEG coated sulfur particles afford electrical conductivity to the sulfur particles. These factors all contributed to the relative high performance of the sulfur-graphene composite cathode material.

The high specific capacity and good cycling stability make our graphene-sulfur composite a promising potential material for future lithium ion batteries with high energy density. It is worth noting that the graphene-sulfur composite could be coupled with silicon based anode materials for rechargeable batteries with significantly higher energy density than currently possible.[31-33] In such batteries, $Li^+$ could be preloaded into either sulfur or silicon.[11,33]

In summary, we have developed a graphene-sulfur composite material by synthesizing submicron sulfur particles coated with PEG containing surfactants and graphene sheets. We proposed that such an approach could render sulfur particles electrically conducting, allow for entrapment of polysulfide intermediates, and accommodate some of the stress and volume expansion during discharge of sulfur. The graphene-sulfur composite showed high specific capacity with relatively good cycling stability as the cathode for Li-S batteries. In the future, it will be of significant interest to further stabilize the composite materials and couple sulfur based cathode with pre-lithiated silicon based anode for high energy density rechargeable batteries.

**Acknowledgement**

This work is supported partly by ONR, Intel and Department of Energy, Office of Basic Energy Sciences, Division of Materials Sciences and Engineering under contract DE-AC02-76SF0051 through the SLAC National Accelerator Laboratory LDRD project. H.W. and Y.Y. acknowledge support from the Stanford Graduate Fellowship. A.J. acknowledges support from the National Defense Science and Engineering Graduate Fellowship.

**Figures**

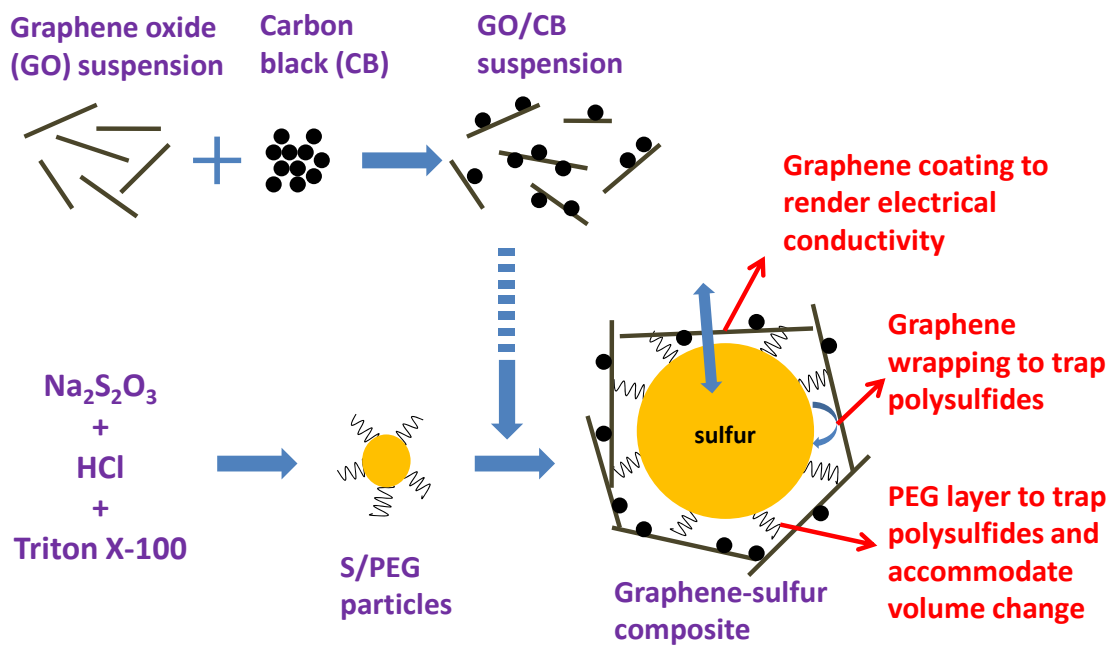

**Figure 1**. Schematic of the synthesis steps for a graphene-sulfur composite, with a proposed schematic structure of the composite.

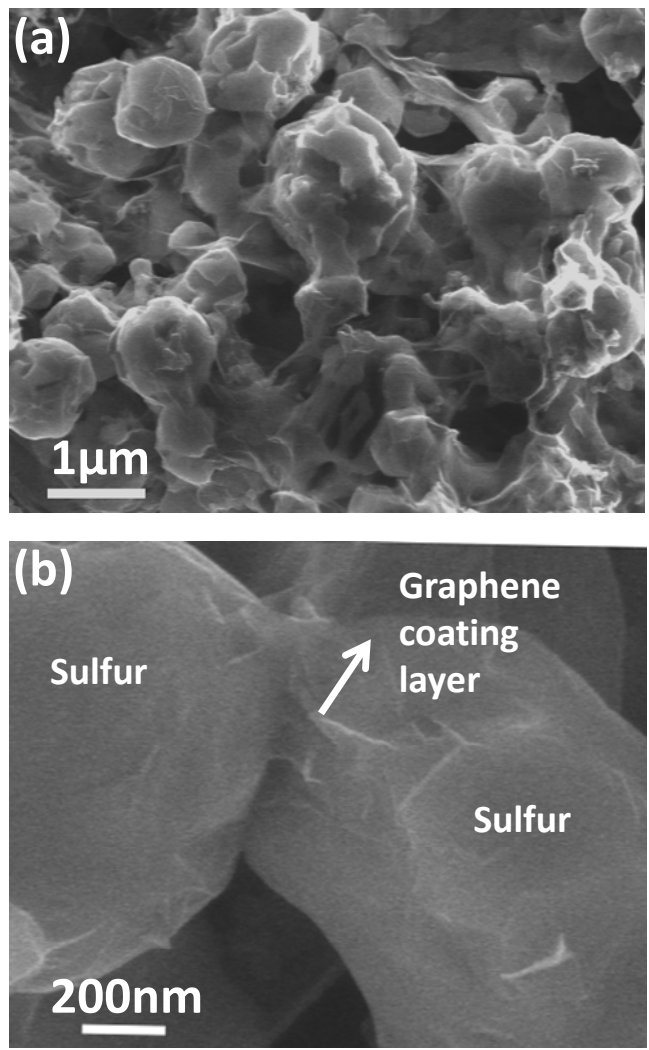

**Figure 2**. SEM characterization of graphene-sulfur composite at low (a) and high (b) magnifications.

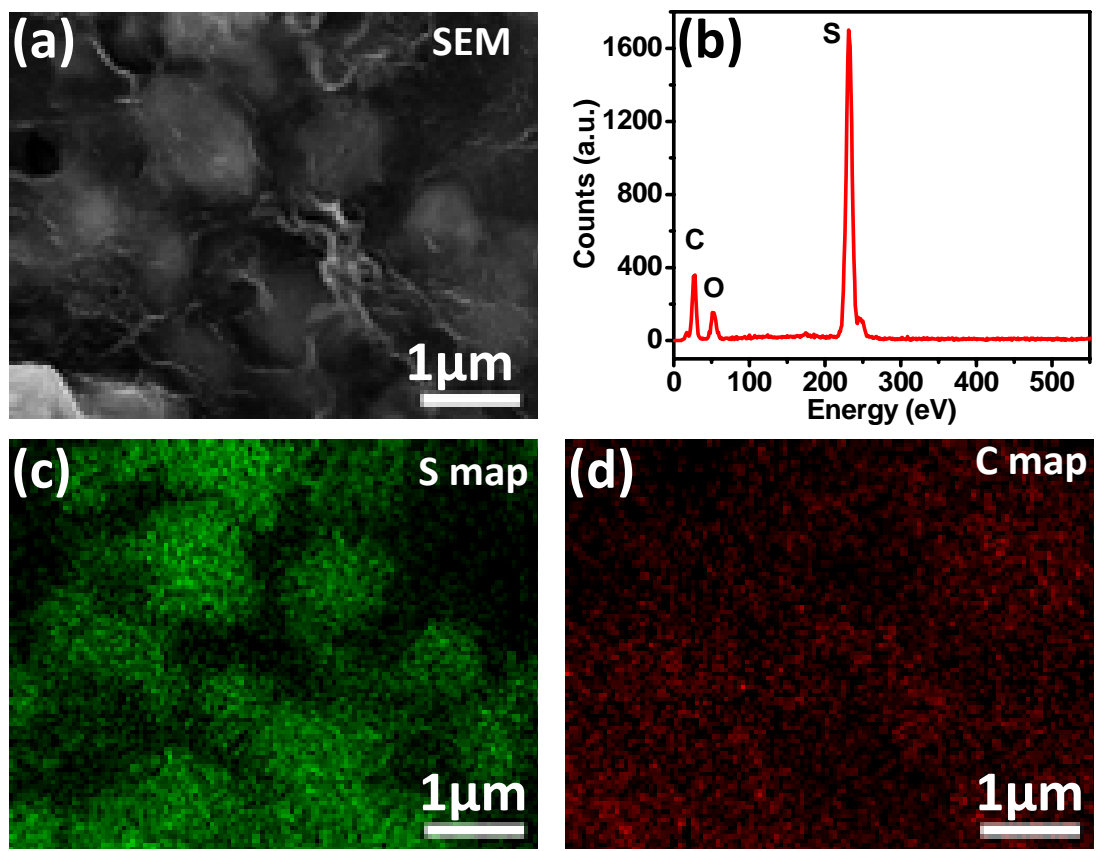

**Figure 3**. Energy dispersive spectroscopic (EDS) characterization of graphene-sulfur composite. (a) SEM image of graphene coated sulfur particles. (b) EDS spectrum captured for the region shown in (a). (c) EDS sulfur mapping of the region shown in (a). (d) EDS carbon mapping of the region shown in (a).

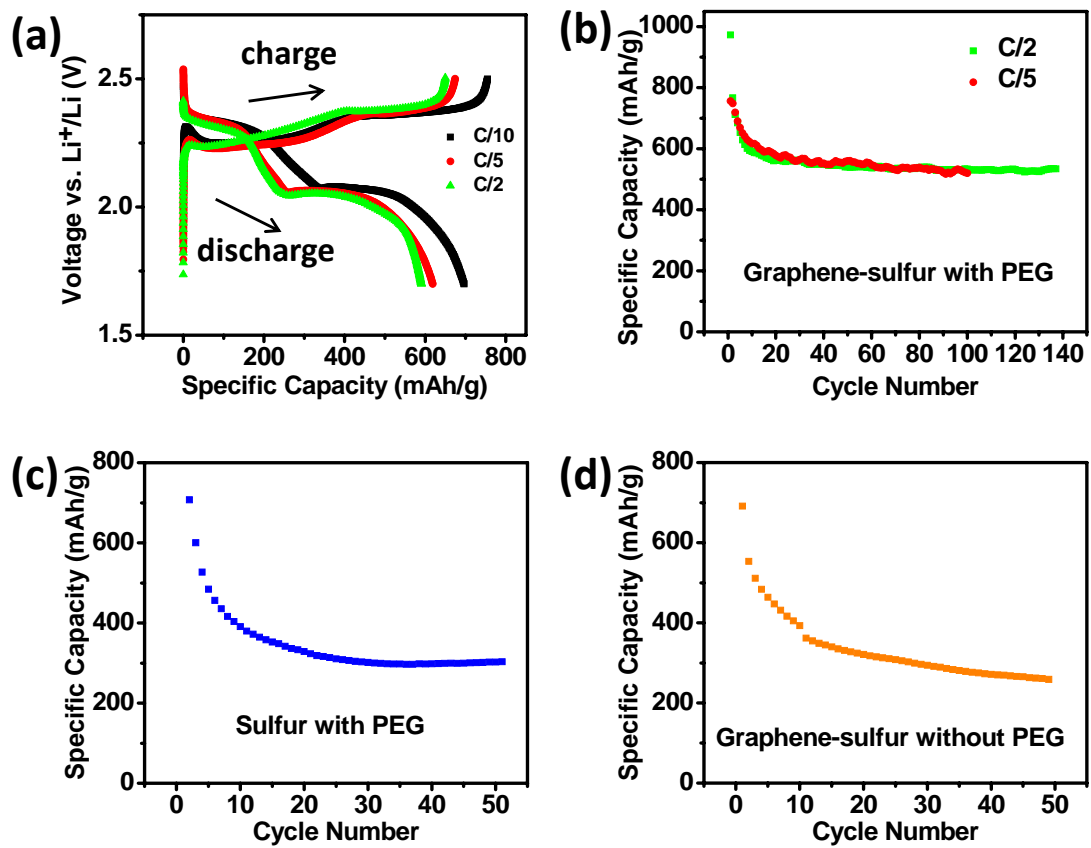

**Figure 4**. Electrochemical characterization of graphene-sulfur composites. (a) 10[th] cycle charge and discharge voltage profiles of the graphene-sulfur composite with PEG coating at various rates. (b) Cycling performance of the same composite as in (a) at rates of ~C/5 and ~C/2. (c) Cycling performance of PEG coated sulfur without graphene coating at the rate of ~C/5. (d) Cycling performance of graphene coated sulfur without any surfactant PEG coating at the rate of ~C/5.